%% file: main.tex
\documentclass[a4paper,fleqn]{cas-dc}

\usepackage[numbers,sort&compress]{natbib}
\usepackage{subcaption}
\usepackage{xurl}
\usepackage{placeins}
\usepackage[textsize=small]{todonotes}
\usepackage{tabularx}
\usepackage{listings}
\usepackage{minted}
\setmintedinline{breaklines,breakanywhere}
\usepackage[most]{tcolorbox}

\definecolor{codebackground}{RGB}{249,249,249}
\definecolor{codeframe}{RGB}{155,155,155}
\definecolor{diffaddition}{RGB}{0,92,55}
\definecolor{diffdeletion}{RGB}{150,38,38}
\definecolor{diffmeta}{RGB}{80,80,80}

\lstdefinestyle{promptstyle}{
    basicstyle=\ttfamily\scriptsize,
    breaklines=true,
    breakatwhitespace=false,
    frame=single,
    columns=fullflexible,
    keepspaces=true,
    showstringspaces=false,
    aboveskip=0.4em,
    belowskip=0.4em,
    xleftmargin=0.5em,
    xrightmargin=0.5em,
    framexleftmargin=0.3em,
    framexrightmargin=0.3em
}

\lstdefinestyle{sourcestyle}{
    basicstyle=\ttfamily\scriptsize,
    breaklines=true,
    breakatwhitespace=false,
    frame=single,
    columns=fullflexible,
    keepspaces=true,
    showstringspaces=false,
    aboveskip=0.5em,
    belowskip=0.5em,
    xleftmargin=0.5em,
    xrightmargin=0.5em,
    framexleftmargin=0.3em,
    framexrightmargin=0.3em,
    numbers=none,
    tabsize=4
}

\lstdefinelanguage{diff}{
    morecomment=[f][\color{diffmeta}\bfseries]{@@},
    morecomment=[f][\color{diffaddition}]{+},
    morecomment=[f][\color{diffdeletion}]{-}
}

\lstdefinestyle{casestyle}{
    style=sourcestyle,
    basicstyle=\ttfamily\fontsize{8.0}{9.4}\selectfont,
    commentstyle=\ttfamily\upshape,
    backgroundcolor=\color{codebackground},
    rulecolor=\color{codeframe},
    framerule=0.45pt,
    framesep=4pt,
    xleftmargin=0.2em,
    xrightmargin=0.2em,
    framexleftmargin=0pt,
    framexrightmargin=0pt,
    aboveskip=0.7em,
    belowskip=0.7em,
    captionpos=b,
    abovecaptionskip=0.45em,
    belowcaptionskip=0pt,
    upquote=true
}

\lstdefinestyle{diffstyle}{
    style=casestyle,
    language=diff
}

\definecolor{findingbg}{RGB}{247,236,204}
\definecolor{findingborder}{RGB}{125,125,125}

\newcounter{finding}
\newenvironment{finding}
{
    \refstepcounter{finding}
    \begin{tcolorbox}[
        colback=findingbg,
        colframe=findingborder,
        boxrule=0.7pt,
        arc=3pt,
        left=8pt,
        right=8pt,
        top=5pt,
        bottom=5pt,
        before skip=6pt,
        after skip=6pt,
        boxsep=0pt
    ]
    \textbf{Finding~\thefinding:}\ 
}
{
    \end{tcolorbox}
}

\def\tsc#1{\csdef{#1}{\textsc{\lowercase{#1}}\xspace}}
\tsc{WGM}
\tsc{QE}
\begin{document}
% Keep hyperref bookmark writing enabled.
% \let\WriteBookmarks\relax
\def\floatpagepagefraction{1}
\def\textpagefraction{.001}

\shorttitle{Evolution of Deprecated APIs and Their Replacements in Python Libraries}    

\shortauthors{He and Xiao}  

\title [mode = title]{Evolution of Deprecated APIs and Their Replacements in Python Libraries}  




\author[1,2]{Gangqiang He}[orcid=0009-0002-7614-1662]

% \cormark[1]


\ead{gangqiang.he@nuaa.edu.cn}



\credit{Conceptualization, Methodology, Data Curation, Software, Writing – Original Draft, Visualization, Investigation, Writing – Review \& Editing, Validation}

\affiliation[1]{organization={College of Computer Science and Technology, Nanjing University of Aeronautics and Astronautics},
            city={Nanjing},
%          citysep={}, % Uncomment if no comma needed between city and postcode
            country={China}}

\affiliation[2]{organization={Key Laboratory for Safety-critical Software Development and Verification, Nanjing University of Aeronautics and Astronautics},
            city={Nanjing},
%          citysep={}, % Uncomment if no comma needed between city and postcode
            country={China}}

\author[1,2,3]{Guanping Xiao}[orcid=0000-0002-9419-4058]
\cormark[1]

% Footnote of the second author

% Email id of the second author
\ead{gpxiao@nuaa.edu.cn}


\credit{Conceptualization, Methodology, Data Curation, Writing – Original Draft, Visualization, Investigation, Writing – Review \& Editing, Validation, Supervision, Funding Acquisition}

\affiliation[3]{organization={State Key Laboratory for Novel Software Technology, Nanjing University},
            city={Nanjing},
%          citysep={}, % Uncomment if no comma needed between city and postcode
            country={China}}

\cortext[1]{Corresponding author}



\iffalse
\begin{abstract}
Here goes the abstract \nocite{*}%% Remove this line from your manuscript.
\end{abstract}
\fi

\begin{abstract}
\relax
\textbf{Context:}
API deprecation is common in library evolution, but migration requires finding replacements and adapting calls. Prior studies have examined deprecation, replacement identification, and client migration separately, leaving deprecated API--replacement API relationships across granularities, releases, and lifecycle states insufficiently understood.

\textbf{Objective:}
We examine replacement locality and parameter-interface differences across API granularities, cross-version changes in similarity-based replacement rankings, and post-deprecation lifecycles from source-definition and original-invocation perspectives.

\textbf{Method:}
We construct 830 maintainer-specified mappings from 33 Python libraries, covering classes, functions, and methods. We compare definitions, track rankings under two migration scenarios using applicable token-, tree-, and graph-based similarity methods, and identify lifecycle events through source inspection and release-by-release execution.

\textbf{Results:}
Replacement locality increases from classes to methods: same-module replacements account for 39.1\%, 55.6\%, and 79.9\% of class, function, and method mappings, respectively. Most mappings have no parameter changes, but function mappings show more frequent and diverse changes. Replacement rankings vary across releases and code representations, and the applicable methods agree on the overall trend in only about 60\% of cases. Other-candidate effects are dominant or mixed in 49.0\% of key versions. After the deprecation announcement, 56.4\% of mappings retain the source definition and 68.6\% retain an executable original invocation. Among 753 mappings with at least one observed ending event, 25.2\% are source-first or invocation-first rather than same-release. Of these cases, 77.4\% are source-first.

\textbf{Conclusions:}
Deprecated API--replacement API relationships depend on API granularity, release context, code representation, candidate competition, and lifecycle state. These results support evolution-aware replacement API recommendation and automated API migration.
\end{abstract}


% \begin{highlights}
% \item 
% \item 
% \item 
% \end{highlights}

%\nocite{*}

% Keywords

\maketitle

% Main text
\input{sec1-introduction}

\input{sec2-study_design}

\input{sec3-rq1}
\input{sec4-rq2}

\input{sec5-rq3}

\input{sec6-implications}

\input{sec7-threats}

\input{sec8-related_work}

\input{sec9-conclusion}

\printcredits

\section*{Declaration of generative AI and AI-assisted technologies in the manuscript preparation process}

During the preparation of this work, the authors used ChatGPT (OpenAI) to assist with implementing parts of the experimental pipeline and analysis scripts, and to improve the language, clarity, and readability of the manuscript. After using this tool, the authors reviewed and edited the content as needed and take full responsibility for the content of the published article.

\section*{Declaration of competing interest}
The authors declare that they have no known competing financial interests or personal relationships that could have appeared to influence the work reported in this paper.

\section*{Acknowledgments}
This work was partially supported by the Open Research Fund of State Key Laboratory of Novel Software Technology under Grant No. KFKT2025B13.

\section*{Data Availability}
The replication package and the dataset are publicly available at \url{https://github.com/PCART-tools/PCBench-D}.

% Template example content retained but disabled.
\iffalse
\section{}\label{}

% Use the style of numbering in square brackets.
% If nothing is used, default style will be taken.
%\begin{enumerate}[a)]
%\item 
%\item 
%\item 
%\end{enumerate}  

%\begin{itemize}
%\item 
%\item 
%\item 
%\end{itemize}  

%\begin{description}
%\item[]
%\item[] 
%\item[] 
%\end{description}  

\clearpage %%Remove this from your manuscript

% Figure
\begin{figure}%[]
  \centering
%    \includegraphics{}
    \caption{}\label{fig1}
\end{figure}

\begin{table}%[]
\caption{}\label{tbl1}
\begin{tabular*}{\tblwidth}{@{}LL@{}}
\toprule
  &  \\ % Table header row
\midrule
 & \\
 & \\
 & \\
 & \\
\bottomrule
\end{tabular*}
\end{table}

% Uncomment and use as the case may be
%\begin{theorem} 
%\end{theorem}

% Uncomment and use as the case may be
%\begin{lemma} 
%\end{lemma}


\section{}\label{}

\printcredits

\bibliographystyle{cas-model2-names}

\bibliography{cas-refs}
\fi

%% Bibliography for the manuscript.
%\bibliographystyle{cas-model2-names}
\renewcommand{\bibfont}{\fontsize{8pt}{9.5pt}\selectfont}
\bibliographystyle{model1-num-names}
\bibliography{cas-refs}



\end{document}

%% file: sec1-introduction.tex
\section{Introduction}

Python third-party libraries evolve continuously, and so do their public APIs.
A deprecation notice informs users that an API should no longer be used and often recommends a replacement~\cite{wang2020python-deprecation,brito2018api-deprecation-replacement}.
Developers must then identify the replacement, compare the two interfaces, and update affected calls to keep their code compatible with later library versions.

Prior work on deprecated API migration has studied API evolution and deprecation practices~\cite{wang2020python-deprecation,zhang2020python-api-evolution}, breaking-change and deprecation detection~\cite{du2022aexpy,vadlamani2021deprecated-api-detection}, replacement API identification and recommendation~\cite{zhu2021relancer,ni2021soar,huang2021repfinder,huang2024python-api-mapping,venkatakrishnan2026libshift}, and invocation adaptation~\cite{zhu2021relancer,ni2021soar,haryono2021mlcatchup,zhang2026pcart,zhu2026deprecated-api-llm,islam2025python-llm-migration,kang2026pig}.
Code similarity based on API signatures, implementations, or program structures is often used to generate or rank replacement candidates~\cite{huang2024python-api-mapping,wu2010aura,huang2021repfinder,venkatakrishnan2026libshift}.
Structural matching has also been used to relate program entities across versions~\cite{dig2006refactoring-detection,kim2007structural-matching}.

A deprecated API--replacement API relationship is more than a static mapping from an old API to a new one.
First, developers and migration tools must identify the replacement among available candidates and then adapt existing calls when the two APIs have different interfaces.
Second, dependency upgrades often span several releases rather than advancing one version at a time.
During such upgrades, the deprecated API, its replacement, and other candidates may all change. 
These changes affect both the code evidence used to identify the replacement and its rank among the candidates.
Thus, the relationship depends on API granularity, source-definition location, parameter interface, and version.

Python supports compatibility mechanisms such as aliases, re-exports, inheritance, and dynamic attribute resolution~\cite{zhang2020python-api-evolution,du2022aexpy,levkivskyi2017dynamic-attribute-deprecation,python2026module-public-api}.
As a result, source-definition removal and original-invocation failure may occur in different versions.
An original invocation may remain executable after source-definition removal because a compatibility mechanism preserves the old access path.
Conversely, the original invocation may fail while the source definition remains because the access path, interface, or implementation has changed.

\begin{figure}[pos=t]
    \centering
    \begin{tikzpicture}[
        >=stealth,
        event/.style={
            draw=black!65,
            rounded corners=1.5pt,
            align=center,
            text width=2.90cm,
            minimum height=1.10cm,
            inner sep=3pt,
            fill=black!4,
            font=\footnotesize,
            text=black
        },
        state/.style={event, dashed, fill=white},
        subcap/.style={align=center, font=\footnotesize, text=black},
        api/.style={anchor=west, align=left, font=\ttfamily\footnotesize, text=black},
        track/.style={anchor=east, align=right, font=\footnotesize, text=black},
        period/.style={align=center, font=\footnotesize, text=black}
    ]
        \node[track] at (1.45,6.15) {Source\\Definition};
        \draw[->, thick] (1.60,6.15) -- (7.65,6.15);
        \fill (3.20,6.15) circle (1.6pt);
        \fill (6.00,6.15) circle (1.6pt);
        \node[align=center,font=\footnotesize,above=2pt,text=black] at (3.20,6.15) {Deprecation\\Announcement $D$};
        \node[align=center,font=\footnotesize,above=2pt,text=black] at (6.00,6.15) {Source-Definition\\Removal $S$};
        \node[period] at (2.40,5.45) {Pre-\\Deprecation\\Period};
        \node[period] at (4.60,5.45) {Source-Definition\\Transition\\Period};
        \node[period] at (6.82,5.45) {Post-\\Removal\\Period};

        \node[track] at (1.45,4.05) {Original\\Invocation};
        \draw[->, thick] (1.60,4.05) -- (7.65,4.05);
        \fill (3.20,4.05) circle (1.6pt);
        \fill (6.00,4.05) circle (1.6pt);
        \draw[dashed,black!55] (3.20,6.00) -- (3.20,4.20);
        \node[align=center,font=\footnotesize,above=2pt,text=black] at (6.00,4.05) {Original-Invocation\\Failure $F$};
        \node[period] at (2.40,3.35) {Pre-\\Deprecation\\Period};
        \node[period] at (4.60,3.35) {Original-Invocation\\Transition\\Period};
        \node[period] at (6.82,3.35) {Post-\\Failure\\Period};
        \node[subcap] at (3.80,2.65) {(a) Lifecycle Events and Periods};

        \node[api] at (0,2.10) {scipy.stats.stats.median\_absolute\_deviation};
        \node[event] (scipy-removal) at (1.75,1.15) {Source-Definition\\Removal\\1.8.0};
        \node[state] (scipy-call) at (5.75,1.15) {Original Invocation\\Remains Executable\\via \texttt{\_\_getattr\_\_}};
        \draw[->, thick] (scipy-removal.east) -- (scipy-call.west);
        \node[subcap] at (3.80,0.35) {(b) Source-first: SciPy};

        \node[api] at (0,-0.30) {Word2Vec.load\_word2vec\_format};
        \node[event] (gensim-failure) at (1.75,-1.25) {Original-Invocation\\Failure\\1.0.0};
        \node[event] (gensim-removal) at (5.75,-1.25) {Source-Definition\\Removal\\4.0.0};
        \draw[->, thick] (gensim-failure.east) -- (gensim-removal.west);
        \node[subcap] at (3.80,-2.05) {(c) Invocation-first: Gensim};
    \end{tikzpicture}
\caption{Lifecycle events, periods, and two cases with different event timing. (a) defines the release periods used in this study. In SciPy (b), source-definition removal occurs in 1.8.0 while \texttt{\_\_getattr\_\_} preserves the original invocation. In Gensim (c), original-invocation failure occurs in 1.0.0 and source-definition removal in 4.0.0.}
    \label{fig:lifecycle-events}
\end{figure}
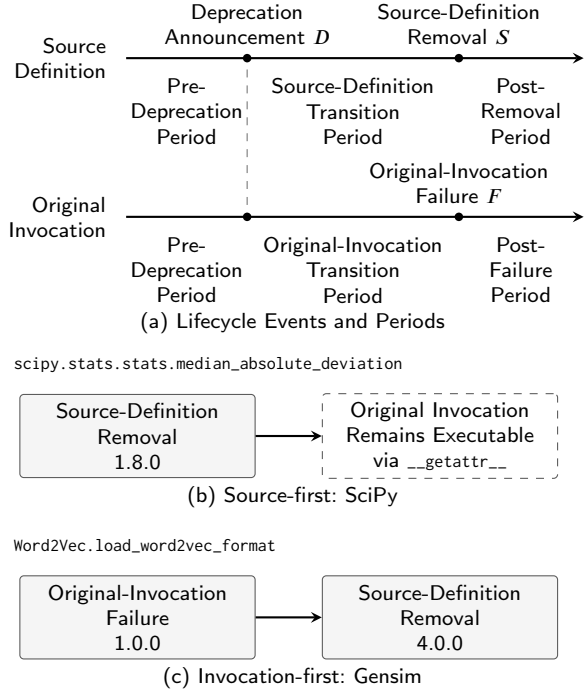

We distinguish three lifecycle events. A \textit{deprecation announcement} is the first explicit notice that an API is deprecated.
\textit{Source-definition removal} is the first release in which the original source definition no longer exists at its recorded definition path.
\textit{Original-invocation failure} is the first observed release in which a validated original invocation no longer runs successfully because of a change to the API, its access path, or its behavior.
Let $D$, $S$, and $F$ denote the releases of these three events, respectively. The \textit{pre-deprecation period} contains releases before $D$. The \textit{source-definition transition period} extends from $D$ through the release immediately preceding $S$, and the \textit{post-removal period} begins at $S$. Similarly, the \textit{original-invocation transition period} extends from $D$ through the release immediately preceding $F$.
Figure~\ref{fig:lifecycle-events} defines these events and periods and illustrates two real-world examples in which the ending events can diverge.

Existing studies usually examine API deprecation, replacement identification, and invocation adaptation separately, leaving the evolution of deprecated API--replacement API relationships insufficiently understood. This leaves three open questions.
First, patterns of replacement locality and parameter-interface differences across API granularities are not yet clear.
These patterns affect the candidate search scope and the adaptation of existing calls.
Second, existing code-similarity-based approaches evaluate API representations at selected version pairs~\cite{huang2024python-api-mapping,wu2010aura,huang2021repfinder,xing2007diff-catchup,venkatakrishnan2026libshift}, although all APIs in the candidate set may change across releases.
The effect of these changes on replacement rankings is not yet clear.
Third, source-definition removal and original-invocation failure may occur in different versions, but their timing and causes have not been systematically studied.

To address these gaps, we examine deprecated API--replacement API relationships in Python third-party libraries.
We construct and verify 830 mappings from 33 libraries, including 207 class-level, 239 function-level, and 384 method-level mappings.
For each mapping, we use changelogs, source-code inspection, and version-by-version execution of a manually validated original invocation to track the three lifecycle events, recording ending events as unobserved when necessary.
We analyze replacement locality and parameter-interface differences, the effect of version evolution on replacement recommendation, and post-deprecation lifecycles.
We address the following three research questions:

\begin{itemize}
\item \textbf{RQ1: How do replacement locality and parameter-interface differences between deprecated APIs and their replacements vary across API granularities?}

\item \textbf{RQ2: How do the rankings of maintainer-specified replacements in code-similarity-based candidate lists change across library releases?}

\item \textbf{RQ3: How do source-definition removal and original-invocation failure relate during the post-deprecation lifecycle, and what mechanisms explain their timing differences?}

\end{itemize}

Our results reveal three main findings. First, replacement locations become more local from classes to methods. Most mappings show no parameter changes, but function-level mappings have more frequent and diverse interface changes. Second, replacement rankings vary across versions and similarity methods, with key changes occurring in different lifecycle periods under the two migration scenarios. Other-candidate effects appear in 49.0\% of the identified key versions, either as the dominant source or as part of a mixed attribution. Third, deprecated APIs commonly have post-deprecation transition periods. Among mappings with at least one observed ending event, 25.2\% are source-first or invocation-first rather than same-release. Of these cases, 77.4\% are source-first, mainly because aliases, re-exports, and inheritance preserve the original invocation. Together, these findings show that deprecated API--replacement API relationships depend on API granularity, version evolution, candidate competition, and lifecycle state.

Our main contributions are as follows:

\begin{itemize}

\item \textbf{Deprecated API--replacement API relationships.} To the best of our knowledge, this is the first large-scale empirical study of maintainer-specified deprecated API--replacement API relationships in Python libraries that jointly examines API granularity, cross-release ranking evolution, and post-deprecation lifecycle states.

\item \textbf{Post-deprecation lifecycle.} We distinguish source-definition removal from original-invocation failure and characterize their transition periods, temporal relationships, and underlying mechanisms. This dual perspective reveals how original invocations can remain executable after source-definition removal or fail before it.

\item \textbf{Dataset.} We provide a curated dataset of 830 deprecated API--replacement API mappings from 33 Python libraries, together with changelog evidence, source definitions, lifecycle-event versions, and validated original invocations. The dataset supports research on API deprecation, replacement recommendation, lifecycle analysis, and automated migration.

\end{itemize}

%% file: sec2-study_design.tex
\section{Study Design}

\begin{figure*}[t]
    \centering
    \includegraphics[width=\textwidth]{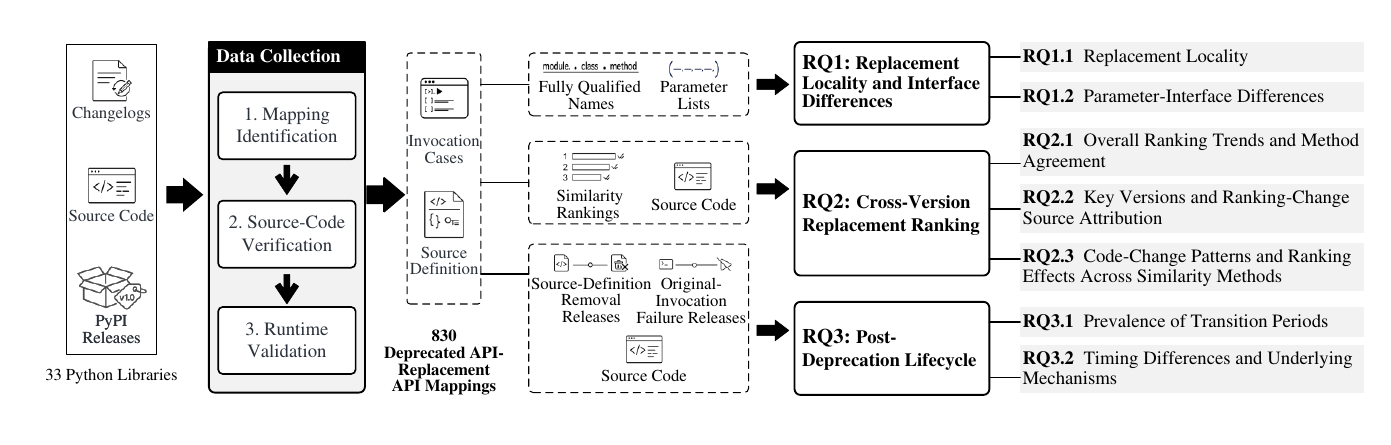}
    \caption{Overview of the dataset construction and research questions.}
    \label{fig:overview}
\end{figure*}

Figure~\ref{fig:overview} summarizes the study. We identify and validate 830 deprecated API--replacement API mappings from 33 Python libraries. RQ1 analyzes their location and parameter interface characteristics, RQ2 analyzes cross-version replacement rankings, and RQ3 compares source-definition removal with original-invocation failure.

\subsection{Dataset Construction}

\subsubsection{Study Scope and Data Sources}

We study 33 Python third-party libraries from a prior benchmark, i.e., PCBench~\citep{zhang2026pcart}. Selected from highly starred GitHub projects, these libraries have widely used callable APIs, documentation, and changelogs and cover several application domains.
We collected all changelogs available by the dataset construction cutoff of March 18, 2025 and considered only relationships reported by that date.

Each mapping must explicitly identify the deprecated API and its replacement in the same-release changelog evidence. Both APIs must belong to the same library, have locatable source definitions, and share the class, function, or method granularity. The deprecated API must also support a validated original invocation across releases. Records outside this scope or without the required evidence are excluded.

\subsubsection{Pilot Study for Changelog Retrieval}

Because deprecation information appears both in dedicated sections and elsewhere in changelogs, a pilot study identified common section names and retrieval keywords.

We randomly sampled five releases per library, yielding 165 changelogs. Two authors independently screened every entry under a broad pilot scope, identified dedicated sections and deprecation-related keywords, and resolved disagreements before merging the retained keywords.

Only 38/165 sampled releases had dedicated deprecation-related sections. Formal collection therefore screens every entry in these sections and uses keyword retrieval elsewhere.
Keyword annotations achieved a Jaccard similarity of 0.852 and Cohen's kappa of 0.907. To cover tense and other word-form variations during retrieval, we manually merged related forms into search stems. The 15 retained stems in Table~\ref{tab:deprecation-keywords} covered 99.46\% of the pilot entries.

\begin{table}[t]
\centering
\caption{Search stems derived from the pilot study.}
\label{tab:deprecation-keywords}

\scriptsize
\setlength{\tabcolsep}{2.5pt}
\renewcommand{\arraystretch}{1.05}

\begin{tabular}{@{}lr lr lr@{}}
\toprule
Stem & Freq.
& Stem & Freq.
& Stem & Freq. \\
\midrule

deprecat  & 211
& in favo   & 22
& recommend & 4 \\

mov       & 156
& renam     & 14
& delet     & 3 \\

instead   & 96
& longer    & 10
& discourag & 3 \\

replac    & 27
& internal  & 7
& refactor  & 2 \\

privat    & 24
& drop      & 6
& migrat    & 2 \\

\bottomrule
\end{tabular}
\end{table}

\subsubsection{Mapping Identification and Validation}

Formal data collection identifies mappings from changelogs, validates source definitions, and verifies original invocations across releases.

\noindent \textit{\textbf{Step 1: Changelog-Based Mapping Identification.}}
For every collected release, we manually screened all entries in dedicated deprecation-related sections and all keyword matches in other sections.
The entry and its same-release context had to identify both APIs explicitly. Library documentation was used only to resolve API identities, not to infer an unstated replacement.

We excluded non-target entities, changes without a corresponding API definition or specific replacement, and deprecated entities defined only as simple aliases. Unresolved granularities or locations proceeded to source validation.
Repeated reports were merged, using the earliest release that explicitly identified both APIs as the deprecation announcement version.
Two authors independently assessed the entries and resolved disagreements through discussion.

\noindent \textit{\textbf{Step 2: Source-Definition Validation.}}
For both APIs, we collected the granularity, definition location and content, and source link, and we identified the deprecated API's source-definition removal version.

We align changelog versions and Git tags with Python-3-compatible final PyPI releases, excluding prereleases and releases without a reliable correspondence. Source-definition removal is the first release without the deprecated API's explicit definition at its original path.

At the announcement, we record the deprecated API's definition in the final preceding release and the replacement's definition in the announcement release. We then trace the deprecated definition through the release sequence to identify the removal version. At removal, we record the deprecated definition in the preceding release and the replacement definition in the removal release. If the deprecated API's definition was already removed before the announcement, we record it in the release before its removal instead.

We locate definitions using names, paths, enclosing classes, and the \mintinline{python}{class} and \mintinline{python}{def} forms, and check them against the changelog, granularity, and adjacent-release evolution. Library source paths take priority over test code.
We retain only same-granularity mappings with two reliably located definitions.
One author collected this evidence, and a second reviewed uncertain cases until consensus.

\noindent \textit{\textbf{Step 3: Original-Invocation Validation.}}
For each source-validated mapping, we execute a validated original invocation across releases through the lifecycle observation cutoff of August 1, 2025. No mappings reported after the dataset construction cutoff are added. Original-invocation failure is the first release in which an API-related change prevents the validated original invocation from executing. If no such release is observed by the cutoff, the event is recorded as unobserved.

GPT-4o (model version: \texttt{gpt-4o-2024-08-06}) was used to generate an initial invocation from the release and source-definition-based fully qualified names (FQNs). We correct and validate it in the final pre-announcement release, or in the release before removal if the definition was already removed. GPT-4o did not judge validity or event versions. We retain direct class instantiations, function calls, and method calls and exclude APIs normally triggered through other language mechanisms.

We execute the unchanged target-API invocation forward from the release immediately preceding the announcement, or from the release preceding the source-definition removal when the removal occurs before the announcement. Corrections to non-API test setup restart the full sequence. Only failures caused by the API, its original access path, interface, or implementation are recorded; environment failures and warnings are excluded.
Two authors independently assessed the direct invocation form and resolved disagreements through discussion.

\subsubsection{Final Dataset}

The final dataset contains 830 mappings: 207 classes, 239 functions, and 384 methods. Every mapping has an explicit changelog relationship, verified source definitions for APIs of the same granularity in the same library, and a validated original invocation. Table~\ref{tab:dataset-summary} summarizes the dataset by library and API granularity.

\begin{table}[t]
\centering
\caption{Final dataset by library and API granularity.}
\label{tab:dataset-summary}
\scriptsize
\setlength{\tabcolsep}{3.5pt}
\renewcommand{\arraystretch}{0.9}
\begin{tabular}{@{}lrrrr@{}}
\toprule
Library & Class & Function & Method & Total \\
\midrule
requests     & 0  & 0  & 1  & 1   \\
tornado      & 4  & 2  & 5  & 11  \\
plotly       & 1  & 1  & 0  & 2   \\
numpy        & 6  & 17 & 0  & 23  \\
networkx     & 2  & 13 & 0  & 15  \\
sympy        & 14 & 18 & 18 & 50  \\
pandas       & 13 & 17 & 94 & 124 \\
rich         & 1  & 1  & 1  & 3   \\
dask         & 1  & 10 & 6  & 17  \\
click        & 0  & 0  & 2  & 2   \\
tensorflow   & 16 & 19 & 6  & 41  \\
polars       & 1  & 8  & 81 & 90  \\
flask        & 0  & 0  & 1  & 1   \\
fastapi      & 0  & 2  & 0  & 2   \\
loguru       & 0  & 0  & 2  & 2   \\
pydantic     & 0  & 1  & 15 & 16  \\
httpx        & 3  & 0  & 24 & 27  \\
lightgbm     & 0  & 1  & 0  & 1   \\
scipy        & 0  & 31 & 0  & 31  \\
transformers & 4  & 2  & 20 & 26  \\
spacy        & 0  & 2  & 1  & 3   \\
pillow       & 0  & 2  & 5  & 7   \\
django       & 25 & 10 & 32 & 67  \\
gensim       & 0  & 0  & 3  & 3   \\
sklearn      & 12 & 9  & 1  & 22  \\
xgboost      & 1  & 0  & 0  & 1   \\
jax          & 13 & 36 & 2  & 51  \\
matplotlib   & 47 & 14 & 48 & 109 \\
keras        & 7  & 0  & 0  & 7   \\
faker        & 0  & 0  & 1  & 1   \\
redis        & 0  & 0  & 3  & 3   \\
torch        & 28 & 23 & 1  & 52  \\
aiohttp      & 8  & 0  & 11 & 19  \\
\midrule
Total        & 207 & 239 & 384 & 830 \\
\bottomrule
\end{tabular}
\end{table}

For each mapping, the dataset records the library, deprecation announcement version, changelog entry, API granularity, and source-definition-based FQNs. It also records the definition locations and contents collected around the deprecation announcement and source-definition removal, the source-definition removal version, the validated original invocation, and the original-invocation failure version or its absence by the observation cutoff.

\subsection{RQ1: Replacement Locality and Interface Differences}
\label{sec2_rq1}

RQ1 examines two characteristics of deprecated API--replacement API relationships. First, it explores where maintainer-specified replacements are located relative to deprecated APIs to inform candidate search scopes. Second, after a replacement has been identified, it examines parameter-interface differences that may affect invocation adaptation.

\noindent \textit{\textbf{RQ1.1: Replacement Locality.}}
Using the deprecated API in the final pre-announcement release and the replacement in the announcement release, we compare source-definition-based FQNs. The comparison covers module paths and names and, for methods, enclosing classes.

Class- and function-level mappings are classified as (1)~\textit{identical}, (2)~\textit{same name, different module}, (3)~\textit{different name, same module}, or (4)~\textit{different name and module}.
Method-level mappings are classified as (1)~\textit{same module and class, different method name}, (2)~\textit{same module, different class, same method name}, (3)~\textit{same module, different class and method name}, (4)~\textit{different module, same method name}, or (5)~\textit{different module and method name}.
We report the count and proportion of each category by API granularity.

\noindent \textit{\textbf{RQ1.2: Parameter-Interface Differences.}}
We compare parameter lists in the same releases used for locality. Function and method parameters come from their definitions. Class parameters come from a custom metaclass's \mintinline{python}{__call__}, if present, or the first available \mintinline{python}{__new__} or \mintinline{python}{__init__}; a class's own \mintinline{python}{__call__} describes instance invocation and is excluded.

We exclude 99 mappings with ambiguous parameter correspondence. For example, the change from \mintinline{python}{glm(data, para)} to \mintinline{python}{ttest_ind(a, b, axis=0, equal_var=True)} may represent renaming plus addition or removal plus addition, which parameter-level evidence alone cannot distinguish reliably.
We analyze the remaining mappings with PCART~\citep{zhang2026pcart}, a tool for Python API parameter changes that identifies additions, removals, renamings, type and position changes, and conversions between positional and keyword arguments.

By granularity, we report identical parameter lists and one, two, three, or at least four change types. Individual types are reported only for mappings with exactly one type.

\subsection{RQ2: Cross-Version Replacement Ranking}
\label{sec2_rq2}

\begin{table*}[t]
\centering
\caption{Code-similarity methods used in RQ2.}
\label{tab:similarity-methods}
\small
\begin{tabularx}{\textwidth}{@{}l>{\raggedright\arraybackslash}X>{\raggedright\arraybackslash}Xl@{}}
\toprule
Method & Representation and preprocessing & Similarity measure & Granularity \\
\midrule

Token & Structured AST-token multiset. API names, parameter types, and return types receive weights 3, 4, and 2, respectively, and other tokens receive weight 1 & Weighted Jaccard & Class, function, method \\

Tree & Normalized AST with consistent renaming of local variables and parameters & Tree-edit similarity & Function, method \\

Graph & Statement-level intraprocedural PDG. Classes use summaries of names, fields, method PDGs, and method categories & Approximate graph-edit similarity with greedy matching & Class, function, method \\
\bottomrule
\end{tabularx}
\end{table*}

Version evolution can change both the code evidence used for replacement recommendation and the set of competing candidates. Using the maintainer-specified replacement as the evaluation target, RQ2 examines how these changes affect its rank as the deprecated API, replacement API, and other candidates evolve across releases. RQ2.1 examines trends and method agreement. RQ2.2 identifies large ranking changes and their source attribution. RQ2.3 examines the associated code-change patterns and their effects across similarity methods.

\subsubsection{Experimental Setup}

\noindent \textit{\textbf{Candidate Scope and Migration Scenarios.}}
Following RQ1, a mapping is retained only if its replacement lies within the parent subpackage of the deprecated API's defining module, a middle ground between a single module and the full library. We then enumerate same-granularity candidates from the parent subpackage of the deeper of the deprecated and replacement FQNs; for \mintinline{python}{lib.pkg.module.dep_api}, these are the candidates under \mintinline{python}{lib.pkg} in each release. This scope contains 153/207 class, 190/239 function, and 380/384 method replacements, yielding 723 mappings for scenario construction.

To distinguish the effects of target-release evolution from those of initial-release variation, we construct two migration scenarios for mappings with an observed source-definition removal. Mappings whose source definition remains at the observation cutoff are excluded from both scenarios.

(1) In the \textit{fixed-deprecated-API scenario} (\texttt{D-fixed}), the deprecated API remains in the final pre-announcement release, or in the release before removal if its definition was already removed, while candidates advance from the announcement through the tenth post-removal release, or the cutoff when fewer exist. This fixes the initial release and varies the target release.

(2) In the \textit{fixed-replacement-API scenario} (\texttt{R-fixed}), candidates remain in the removal release, while the deprecated API advances from the tenth pre-announcement release, or the earliest available release, through the final pre-removal release. This varies the initial release and fixes the target release.

Release points without the evolving API's required definition are excluded.

\noindent \textit{\textbf{Code-Similarity Computation and Ranking.}}
We select three commonly used code-similarity methods~\cite{zakeri-nasrabadi2023code-similarity-review}, i.e., token-based, tree-based, and graph-based, to represent code at different levels. Table~\ref{tab:similarity-methods} summarizes these methods. We implemented all three similarity methods for this study. Their source code, parameter settings, and execution environments are provided in our replication package.

After computing similarity, each method removes the zero-similarity release points at the two ends of its own sequence, where the deprecated and replacement APIs share no code elements and the replacement's rank is therefore meaningless---determined solely by the candidate-set size rather than its competitiveness---so retaining them would distort the Mann--Kendall test. These zero points occur only at the sequence ends, leaving the remaining releases consecutive. If, after this removal, any method keeps fewer than ten release points, the migration scenario is discarded to support reliable trend analysis; this yields 529 \texttt{D-fixed} and 443 \texttt{R-fixed} scenarios, covering 557 mappings.

For each release and method, candidates are sorted by decreasing similarity. One-based competition ranking assigns equal scores the same rank and skips subsequent tied positions, producing the replacement-rank sequences used below.

\subsubsection{RQ2.1: Overall Ranking Trends and Method Agreement}

For each mapping, scenario, and method, we apply the Mann--Kendall test ($\alpha=0.05$) to the transformed replacement-rank sequence used by the implementation~\cite{mann1945nonparametric-trend}. Let $r_t$ denote the raw numerical rank at release $t$, where a smaller value indicates a better position, and let $q_t=-r_t$ denote the transformed value used for trend testing. Let $S$ and $p$ denote the Mann--Kendall statistic and $p$-value computed from $q_t$, respectively. Therefore, $S>0$ with $p<\alpha$ indicates that the raw numerical rank decreases and is classified as \textit{significant ranking improvement}; $S>0$ with $p\geq\alpha$ indicates a \textit{non-significant improvement tendency}; $S<0$ with $p<\alpha$ indicates that the raw numerical rank increases and is classified as \textit{significant ranking decline}; $S<0$ with $p\geq\alpha$ indicates a \textit{non-significant decline tendency}; and $S=0$ indicates \textit{no directional trend}. Repeated rank values are retained as ties. The replication package provides the analysis script and execution environment used for this classification.

For each mapping and scenario, \textit{method agreement} requires a common category from token and graph for classes, and from all three methods for functions and methods. Note that tree-based similarity is not applied at the class level because tree-edit-distance computation was infeasible for several class definitions containing thousands of lines. Class-level comparisons therefore use token- and graph-based similarity throughout.
Other combinations indicate \textit{method disagreement}.

We report agreement and disagreement distributions by scenario, granularity, and source-definition transition-period status. Five categories yield $5^2$ possible method-specific assignments for classes and $5^3$ for functions and methods.

\subsubsection{RQ2.2: Key Versions and Ranking-Change Source Attribution}
\label{sec:rq22-design}

For adjacent valid releases $v_i$ and $v_{i+1}$, let $N_i$ be the candidate count in $v_i$, $\mathcal{S}$ the applicable methods, and $p_m(v)$ the replacement rank. The normalized ranking-change magnitude is

\begin{equation}
q_m(i)=
\frac{\left|p_m(v_{i+1})-p_m(v_i)\right|}{N_i}.
\label{eq:normalized-rank-change}
\end{equation}

The absolute difference captures magnitude rather than direction. We require $N_i\geq10$ so that a one-position shift cannot exceed 10\%, and identify $v_{i+1}$ as a key version when any method reaches the 0.10 threshold:

\begin{equation}
v_{i+1}\in\mathcal{K}
\iff N_i\geq10
\ \land\
\max_{m\in\mathcal{S}}q_m(i)\geq0.10.
\label{eq:key-version}
\end{equation}

For mappings with a source-definition transition period, we compare that period with the post-removal period in \texttt{D-fixed} and with the pre-deprecation period in \texttt{R-fixed}, assigning each key version by $v_{i+1}$. For each period and granularity, we report its share of key versions and the proportion of valid release points identified as key versions.

To distinguish whether a key rank change reflects the deprecated--replacement pair, other candidates, or both, let $\mathcal{S}_i^{\Delta}=\{m\in\mathcal{S}:p_m(v_{i+1})\neq p_m(v_i)\}$ denote the methods whose replacement rank changes across the key-version transition. Methods with unchanged ranks are excluded from attribution; a method remains included when its change is nonzero, even if $q_m(i)<0.10$. For each $m\in\mathcal{S}_i^{\Delta}$, let $R$ be the replacement and $\mathcal{X}_m(i)$ the associated candidates. For improvement, they rank above $R$ in $v_i$ but below or tied with $R$, or are absent, in $v_{i+1}$. An absent candidate or unchanged replacement similarity yields other-candidate dominance. Otherwise, with similarity changes $\Delta s_{m,R}(i)$ and $\Delta s_{m,c}(i)$, define $r_{m,c}(i)=\Delta s_{m,c}(i)/\Delta s_{m,R}(i)$ and

\begin{equation}
a_{m,c}(i)=
\begin{cases}
P, & -1<r_{m,c}(i)<1,\\
C, & \lvert r_{m,c}(i)\rvert>1,\\
M, & r_{m,c}(i)=-1,
\end{cases}
\label{eq:candidate-attribution}
\end{equation}
where $P$, $C$, and $M$ denote the \textit{pair-dominant}, \textit{other-candidate-dominant}, and \textit{mixed} categories, respectively. At the candidate level, $M$ represents equal-magnitude changes in opposite directions. The relative-position condition excludes $r_{m,c}(i)=1$.

Let $n_{m,P}(i)$, $n_{m,C}(i)$, and $n_{m,M}(i)$ be the numbers of candidates in $\mathcal{X}_m(i)$ assigned to the three categories. The method-level attribution category is

\begin{equation}
A_m(i)=
\begin{cases}
P, & n_{m,P}(i)>\max(n_{m,C}(i),n_{m,M}(i)),\\
C, & n_{m,C}(i)>\max(n_{m,P}(i),n_{m,M}(i)),\\
M, & \text{otherwise}.
\end{cases}
\label{eq:method-attribution}
\end{equation}

A tie between $P$ and $C$, or any maximum involving $M$, yields $M$. At the key-version level, $A(i)$ retains the common $A_m(i)$ when all methods in $\mathcal{S}_i^{\Delta}$ agree and yields $M$ otherwise. Thus, $M$ covers equal candidate-level contributions, method-level ties, and disagreement among rank-changing methods.
For ranking decline, we reverse the version direction and apply the same procedure.

Because the candidate set evolves only in \texttt{D-fixed}, whereas the deprecated API evolves in \texttt{R-fixed}, we analyze candidate-set changes only for \texttt{D-fixed}. We report the proportion of key versions for which the absolute candidate-count difference between $v_i$ and $v_{i+1}$ is at least $0.10N_i$ and, by attribution category, the ratio of removed, added, or source-modified candidate instances to candidates in the preceding releases $v_i$. These measures capture net size change and broader candidate-set change, respectively.

\subsubsection{RQ2.3: Code-Change Patterns and Ranking Effects Across Similarity Methods}
\label{sec:rq23-design}

RQ2.3 identifies the main code-change patterns underlying the key-version rank changes. We use block-wise restoration to isolate their dominant change blocks. \texttt{D-fixed} scenarios include only pair-dominant key versions because the other two categories commonly involve many changing candidates. \texttt{R-fixed} scenarios include all three attribution categories.

We split the changed API's diff into contiguous blocks, excluding comment-only changes. The changed API is the replacement in \texttt{D-fixed} scenarios and the deprecated API in \texttt{R-fixed} scenarios.
For each similarity method, we restore one block to its pre-key-version state while keeping the other blocks in their post-key-version states, then recompute the replacement rank.

A \textit{dominant change block} produces the largest reverse rank change when restored: decline after an original improvement, or improvement after an original decline. All ties are retained. If restoring the block that yields the largest reverse change does not move the replacement rank back across its original value, we designate no \textit{dominant change block} for that method, attributing the change instead to the combined effect of multiple blocks. If the original rank is unchanged, every restored block that also leaves it unchanged is retained as a \textit{no-rank-change-block}. These two types form the \textit{analysis change blocks}.
An analysis change block selected by multiple methods within a key version is counted once, with its ranking effect under each method recorded for that unique block.

For each unique block, its inferred effect when introduced is recorded as \textit{improvement}, \textit{decline}, or \textit{unchanged} for every method, yielding $3^3$ ranking-effect combinations. We inductively code the block's code-change pattern.
Blocks are grouped by scenario, key-version attribution category, API granularity, and ranking-effect combination. We analyze patterns occurring at least five times in relation to these characteristics and the code representations.
One author codes all blocks, and a second reviews every label. Disagreements are resolved through discussion.

\begin{table*}[width=.74\textwidth,pos=t]
    \centering
    \caption{Distribution of source-code location relationships between deprecated and replacement APIs.}
    \label{tab:rq11-location}

    \begin{tabular}{lcc}
        \toprule
        \multicolumn{3}{l}{\textbf{Class and function granularities}} \\
        \midrule
        \multicolumn{1}{c}{Source-code location relationship}
            & Class ($N=207$)
            & Function ($N=239$) \\
        \midrule
        Identical
            & 11 (5.3\%)
            & 17 (7.1\%) \\
        Same name, different module
            & 55 (26.6\%)
            & 60 (25.1\%) \\
        Different name, same module
            & 70 (33.8\%)
            & 116 (48.5\%) \\
        Different name and module
            & 71 (34.3\%)
            & 46 (19.2\%) \\
        \midrule
        \multicolumn{3}{l}{\textbf{Method granularity}} \\
        \midrule
        \multicolumn{1}{c}{Source-code location relationship}
            & \multicolumn{2}{c}{Method ($N=384$)} \\
        \midrule
        Same module and class, different method name
            & \multicolumn{2}{c}{291 (75.8\%)} \\
        Same module, different class, same method name
            & \multicolumn{2}{c}{8 (2.1\%)} \\
        Same module, different class and method name
            & \multicolumn{2}{c}{8 (2.1\%)} \\
        Different module, same method name
            & \multicolumn{2}{c}{58 (15.1\%)} \\
        Different module and method name
            & \multicolumn{2}{c}{19 (4.9\%)} \\
        \bottomrule
    \end{tabular}
\end{table*}

\subsection{RQ3: Post-Deprecation Lifecycle}
\label{sec2_rq3}

A deprecation announcement does not determine when an API's original source definition is removed or its original invocation stops executing. These ending events capture code maintenance and user-facing compatibility, respectively, and may occur in different releases. RQ3 quantifies the prevalence of post-deprecation transition periods from both perspectives, compares their ending events, and explains differences in their timing.

We define three events: deprecation announcement, source-definition removal, and original-invocation failure. They mark the first explicit notice, the first release without the API's explicit definition at its original source path, and the first release in which the validated original invocation no longer executes, respectively. As shown in Figure~\ref{fig:lifecycle-events}, their releases are denoted by $D$, $S$, and $F$.

The ranges $D\leq v<S$ and $D\leq v<F$ define the \textit{source-definition transition period} and \textit{original-invocation transition period}, respectively. RQ3.1 measures the prevalence of both periods across API granularities. RQ3.2 compares their ending events and analyzes the intermediate states and mechanisms when their timing differs. RQ3 covers all 830 mappings through the lifecycle observation cutoff of August 1, 2025.

\noindent \textit{\textbf{RQ3.1: Prevalence of Transition Periods.}}
A transition period exists when its ending event occurs after the announcement release or remains unobserved by the lifecycle observation cutoff. An event in the announcement release indicates no corresponding period, while an unobserved ending establishes the period's existence but not its endpoint. We report the proportions of mappings with and without each period.

\begin{figure}[pos=t]
    \centering
    \includegraphics[width=.9\linewidth]{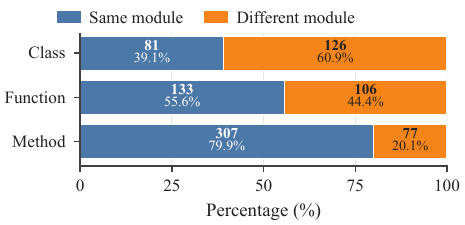}
    \caption{Module locality of replacement APIs across different API granularities.}
    \label{fig:rq11-locality}
\end{figure}

\noindent \textit{\textbf{RQ3.2: Timing Differences and Underlying Mechanisms.}}
We classify each mapping as \textit{same-release} when both ending events occur in the same release, \textit{invocation-first} when original-invocation failure occurs first, \textit{source-first} when source-definition removal occurs first, or \textit{neither observed} when neither occurs by the lifecycle observation cutoff. A one-sided unobserved event is included in the corresponding first-event category, and \textit{same-release} concerns only the ending events. We report all four categories and, among mappings with at least one observed ending event, compare same-release cases with source-first and invocation-first cases.

For invocation-first and source-first mappings, we examine the API state between the events, or from the observed event to the lifecycle observation cutoff when the other remains unobserved. We identify the mechanisms associated with the timing difference using evidence from the implementation, API access path, released-package contents, and runtime exposure. One author assigns initial mechanism codes and groups similar codes into categories. A second author reviews every label, and disagreements are resolved through discussion. We report mechanism distributions separately for invocation-first and source-first mappings.

%% file: sec3-rq1.tex
\section{RQ1: Replacement Locality and Interface Differences}

\subsection{RQ1.1: Replacement Locality}

Table~\ref{tab:rq11-location} reports the location relationships defined in Section~\ref{sec2_rq1}: four categories based on module path and API name for classes and functions, and five categories that also account for the enclosing class for methods.

Class-level replacements are distributed across module and name changes: 70 (33.8\%) have a different name in the same module, 71 (34.3\%) differ in both name and module, and 55 (26.6\%) retain the name but move to another module.
Function-level replacements are more concentrated within the original module: 116 (48.5\%) have a different name in the same module, whereas 46 (19.2\%) differ in both name and module.
Method-level replacements show the strongest locality: 291 (75.8\%) remain in the same module and class under a different method name. Only 16 (4.2\%) move to another class within the same module, and 77 (20.1\%) cross module boundaries.

Figure~\ref{fig:rq11-locality} aggregates these categories by module locality. The \textit{same-module} group contains \textit{identical} and \textit{different name, same module} mappings for classes and functions, and the first three method categories in Table~\ref{tab:rq11-location}.
The proportion of same-module replacements increases from 39.1\% for classes to 55.6\% for functions and 79.9\% for methods. Method locality is also evident below the module level: 75.8\% of method replacements remain in the original class.

\begin{finding}
Replacement locality varies with API granularity. Same-module replacements account for 39.1\% of class, 55.6\% of function, and 79.9\% of method mappings. Among method mappings, 75.8\% of replacements remain in the original class.
\label{finding:1}
\end{finding}

\medskip

\subsection{RQ1.2: Parameter-Interface Differences}

After excluding 99 mappings with ambiguous parameter correspondences, RQ1.2 analyzes 731 mappings: 174 classes, 210 functions, and 347 methods. We first report the number of PCART change types per mapping and then examine the specific types in mappings with exactly one change type.

\begin{figure}[pos=t]
    \centering
    \includegraphics[width=.95\linewidth]{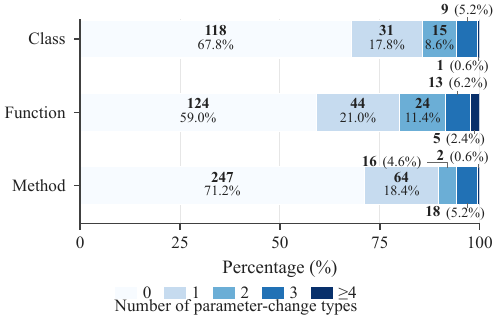}
    \caption{Distribution of parameter change complexity across different API granularities.}
    \label{fig:rq12-complexity}
\end{figure}

\noindent \textit{\textbf{Parameter Change Complexity.}}
Figure~\ref{fig:rq12-complexity} reports the number of parameter change types per mapping. Zero denotes no detected change. The remaining categories denote one, two, three, and at least four change types.

Mappings with no detected parameter changes form the majority at all three granularities: 118/174 classes (67.8\%), 124/210 functions (59.0\%), and 247/347 methods (71.2\%).

Parameter changes are most frequent for functions: 86/210 mappings (41.0\%) contain at least one change type, compared with 32.2\% of classes and 28.8\% of methods. Functions also have the largest proportion of mappings with at least two change types (20.0\%), compared with 14.4\% for classes and 10.4\% for methods.

Parameter-interface differences are therefore more frequent and varied among function replacements, whereas method replacements have the most stable parameter interfaces.

\begin{finding}
Most mappings have no detected parameter changes. Functions show the highest change rates: 41.0\% contain at least one change type and 20.0\% contain at least two.
\label{finding:2}
\end{finding}

\begin{figure}[pos=t]
    \centering
    \includegraphics[width=.9\linewidth]
        {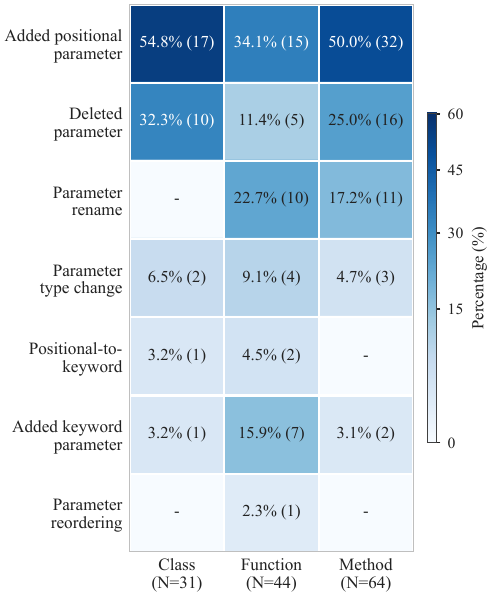}
    \caption{Distribution of single parameter change types across different API granularities.}
    \label{fig:rq12-single-type}
\end{figure}

\noindent \textit{\textbf{Single-Type Parameter Changes.}}
To identify the individual forms of interface difference, we analyze the 139 mappings with exactly one change type: 31 classes, 44 functions, and 64 methods. Figure~\ref{fig:rq12-single-type} reports the count and within-granularity proportion of each type.

Positional-parameter addition is the most frequent single change type for classes (17/31, 54.8\%), functions (15/44, 34.1\%), and methods (32/64, 50.0\%). Overall, it accounts for 64/139 mappings (46.0\%), followed by parameter removal (31, 22.3\%) and renaming (21, 15.1\%).

Class changes are concentrated in positional-parameter addition and removal (27/31, 87.1\%), while method changes are concentrated in positional-parameter addition, removal, and renaming (59/64, 92.2\%).

Function changes are more diverse: positional-parameter addition accounts for 34.1\%, renaming for 22.7\%, keyword-parameter addition for 15.9\%, removal for 11.4\%, and type changes for 9.1\%. All seven single change types occur for functions, whereas some are absent for classes and methods.

These results describe parameter changes in API definitions. Their effects on existing calls also depend on factors such as whether added parameters have default values.

\begin{finding}
Among mappings with one change type, positional-parameter addition is the most frequent type at every granularity and accounts for 46.0\% overall. Function replacements include all seven types and have the most diverse distribution.
\label{finding:3}
\end{finding}

\medskip

%% file: sec4-rq2.tex
\section{RQ2: Cross-Version Replacement Ranking}

\subsection{RQ2.1: Overall Ranking Trends and Method Agreement}

Different code representations may lead to different ranking-trend assessments for the same mapping and migration scenario. We therefore first measure how often the similarity methods assign the same trend category and then examine the distributions of agreed and disagreed assessments.
The analysis covers 557 mappings, yielding 529 valid \texttt{D-fixed} and 443 valid \texttt{R-fixed} scenarios.

\begin{table}[width=\columnwidth,pos=t]
\centering
\caption{Agreement among similarity methods by migration scenario, API granularity, and presence of a source-definition transition period.}
\label{tab:rq21-consistency}
\setlength{\tabcolsep}{3.7pt}
\renewcommand{\arraystretch}{1.02}
\begin{tabular}{@{}llrr@{}}
\toprule
API granularity & Transition period & Valid & Agreement \\
\midrule
\multicolumn{4}{c}{\textbf{\texttt{D-fixed}}} \\
\midrule
\multirow{2}{*}{Class}
& Present & 43  & 27 (62.8\%) \\
& Absent  & 59  & 44 (74.6\%) \\
\addlinespace[1pt]
\multirow{2}{*}{Function}
& Present & 66  & 39 (59.1\%) \\
& Absent  & 50  & 40 (80.0\%) \\
\addlinespace[1pt]
\multirow{2}{*}{Method}
& Present & 167 & 81 (48.5\%) \\
& Absent  & 144 & 91 (63.2\%) \\
\midrule
\textbf{Overall} &  & \textbf{529} & \textbf{322 (60.9\%)} \\
\midrule
\multicolumn{4}{c}{\textbf{\texttt{R-fixed}}} \\
\midrule
\multirow{2}{*}{Class}
& Present & 39  & 25 (64.1\%) \\
& Absent  & 31  & 25 (80.6\%) \\
\addlinespace[1pt]
\multirow{2}{*}{Function}
& Present & 64  & 35 (54.7\%) \\
& Absent  & 52  & 39 (75.0\%) \\
\addlinespace[1pt]
\multirow{2}{*}{Method}
& Present & 158 & 61 (38.6\%) \\
& Absent  & 99  & 81 (81.8\%) \\
\midrule
\textbf{Overall} &  & \textbf{443} & \textbf{266 (60.0\%)} \\
\bottomrule
\end{tabular}
\end{table}

\begin{figure}[pos=t]
    \centering
    \begin{subfigure}[t]{\linewidth}
        \centering
        \includegraphics[width=\linewidth]{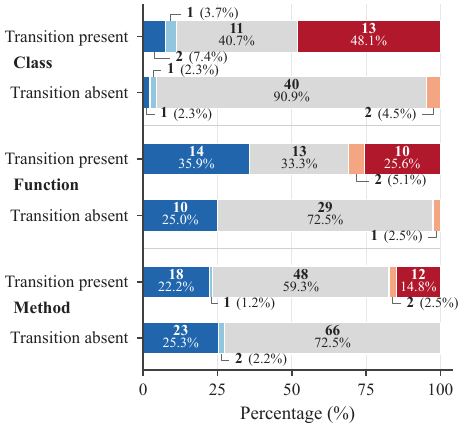}
        \caption{\texttt{D-fixed}}
        \label{fig:rq21-fix-d}
    \end{subfigure}
    \vspace{0.5em}
    \begin{subfigure}[t]{\linewidth}
        \centering
        \includegraphics[width=\linewidth]{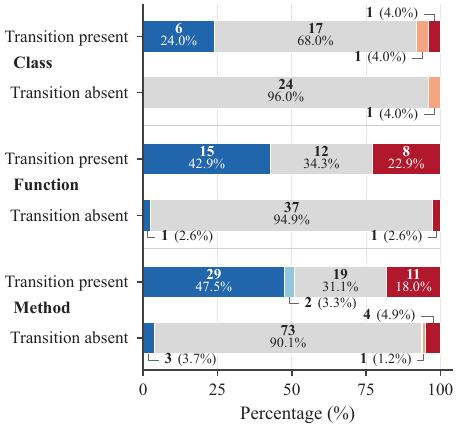}
        \caption{\texttt{R-fixed}}
        \label{fig:rq21-fix-r}
    \end{subfigure}
    \vspace{0.3em}
    \includegraphics[width=\linewidth]{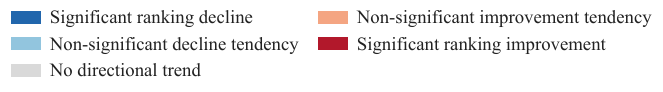}
    \caption{Distribution of replacement API ranking trends among cases with method agreement.}
    \label{fig:rq21-consistent-trends}
\end{figure}

Table~\ref{tab:rq21-consistency} shows similar overall agreement in the two scenarios: 60.9\% (322/529) for \texttt{D-fixed} and 60.0\% (266/443) for \texttt{R-fixed}. Within every API granularity and scenario, agreement is higher when the source-definition transition period is absent. When this period is present, agreement decreases from classes to functions and methods in both scenarios. Because class-level cases use two similarity methods whereas function- and method-level cases use three, these cross-granularity rates describe the observed distributions rather than an isolated granularity effect.

To identify the trends underlying this agreement, Figure~\ref{fig:rq21-consistent-trends} compares the agreed cases. Across both scenarios and all three API granularities, the no-directional-trend segment is consistently larger when the source-definition transition period is absent. Combining cases by period status, 84.1\% (269/320) show no directional trend when the period is absent, compared with 44.8\% (120/268) when it is present. Among cases in which the period is present, 31.3\% (84/268) show significant ranking decline and 20.5\% (55/268) show significant ranking improvement. The remainder show non-significant tendencies.

\begin{figure*}[pos=!t]
    \centering
    \begin{subfigure}[t]{0.405\textwidth}
        \centering
        \includegraphics[width=\linewidth]{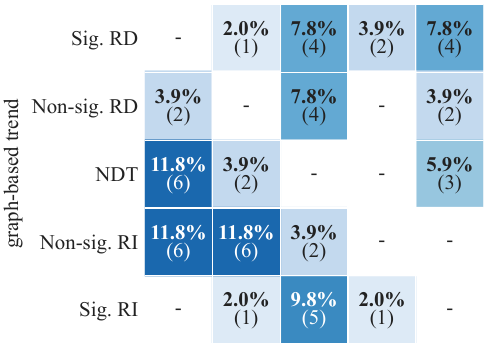}
        \caption{class: graph-by-token}
        \label{fig:rq21-disagreement-class}
    \end{subfigure}
    \hfill
    \begin{subfigure}[t]{0.29\textwidth}
        \centering
        \includegraphics[width=\linewidth]{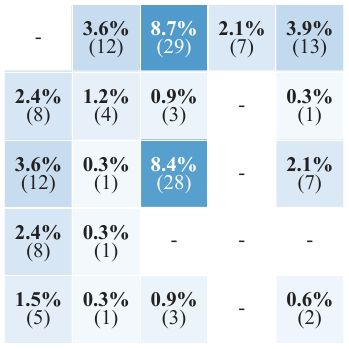}
        \caption{tree-based: Sig. RD}
        \label{fig:rq21-disagreement-tree-sig-dec}
    \end{subfigure}
    \hfill
    \begin{subfigure}[t]{0.29\textwidth}
        \centering
        \includegraphics[width=\linewidth]{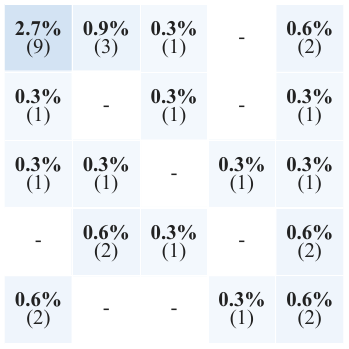}
        \caption{tree-based: Non-sig. RD}
        \label{fig:rq21-disagreement-tree-dec}
    \end{subfigure}

    \vspace{0.6em}

    \begin{subfigure}[t]{0.405\textwidth}
        \centering
        \includegraphics[width=\linewidth]{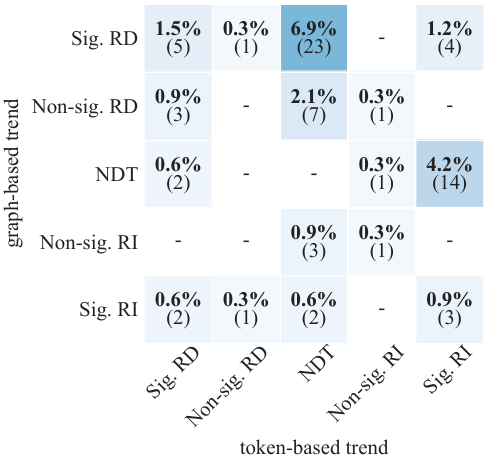}
        \caption{tree-based: NDT}
        \label{fig:rq21-disagreement-tree-none}
    \end{subfigure}
    \hfill
    \begin{subfigure}[t]{0.29\textwidth}
        \centering
        \includegraphics[width=\linewidth]{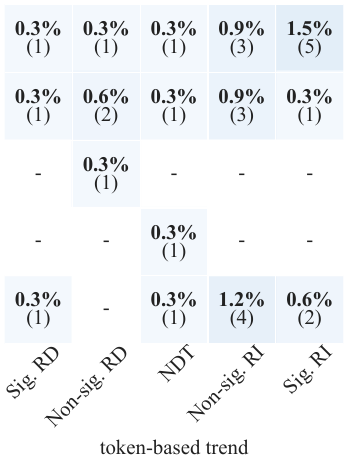}
        \caption{tree-based: Non-sig. RI}
        \label{fig:rq21-disagreement-tree-inc}
    \end{subfigure}
    \hfill
    \begin{subfigure}[t]{0.29\textwidth}
        \centering
        \includegraphics[width=\linewidth]{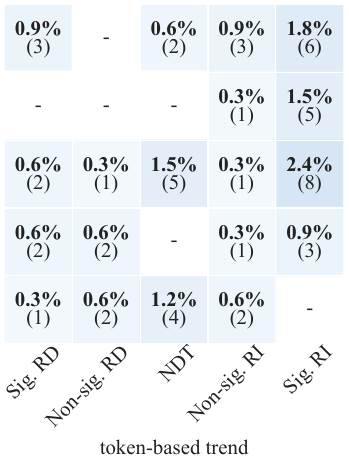}
        \caption{tree-based: Sig. RI}
        \label{fig:rq21-disagreement-tree-sig-inc}
    \end{subfigure}

    \vspace{0.4em}

    \begin{minipage}{0.40\textwidth}
        \centering
        \includegraphics[width=\linewidth]{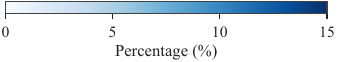}
    \end{minipage}

    \caption{Distribution of trend combinations among cases with method disagreement. (a) shows the distribution of the 51 class-level disagreement cases, whereas (b)--(f) show token-based and graph-based trend combinations for the 333 function- and method-level disagreement cases, conditioned on the five tree-based trend categories. RI, RD, and NDT denote ranking improvement, ranking decline, and no directional trend; Sig. and Non-sig. denote significant and non-significant, respectively.}
    \label{fig:rq21-divergence}
\end{figure*}

We next examine cases with method disagreement. Figure~\ref{fig:rq21-divergence} summarizes the trend-category combinations produced by the applicable similarity methods.
These disagreements mainly concern trend direction, not only statistical significance. When significant and non-significant results in the same direction are treated as one category, 47/51 (92.2\%) class-level cases and 284/333 (85.3\%) function- and method-level cases still show different conclusions about whether the replacement ranking improves, declines, or has no directional trend.

The disagreements also occur across many method combinations. Class-level cases include 16 of the 20 possible disagreement combinations, and the most frequent combination accounts for only 11.8\% (6/51). Function- and method-level cases include 86 of the 120 possible disagreement combinations, and the most frequent accounts for only 8.7\% (29/333). Thus, no single combination dominates the disagreements.

\begin{finding}
Similarity methods agree on the ranking trend in about 60\% of cases. Among agreed cases, mappings for which the source-definition transition period is absent mostly have no directional trend, whereas mappings for which this period is present show a mix of ranking improvement, decline, and no directional trend. Most method disagreements concern trend direction rather than significance alone.
\label{finding:4}
\end{finding}

\subsection{RQ2.2: Key Versions and Ranking-Change Source Attribution}

Overall trends do not show when large ranking changes occur. They also do not show whether the changes arise from the deprecated--replacement pair or the surrounding candidate set. As defined in Section~\ref{sec:rq22-design}, a key version is the later release in an adjacent-release pair when, under at least one similarity method, the replacement rank changes by at least 10\% of the candidate count in the earlier release. Only pairs with at least ten candidates are considered. Using this definition, we identify 306 key versions, with 133 in \texttt{D-fixed} and 173 in \texttt{R-fixed}, and analyze their lifecycle-period distribution and ranking-change source attribution.

\begin{figure}[pos=t]
    \centering
    \begin{subfigure}[t]{.9\linewidth}
        \centering
        \includegraphics[width=\linewidth]{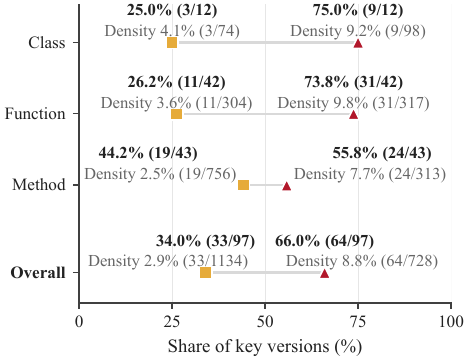}
        \caption{\texttt{D-fixed}}
        \label{fig:rq23-fix-d}
    \end{subfigure}
    \vspace{0.5em}
    \begin{subfigure}[t]{.9\linewidth}
        \centering
        \includegraphics[width=\linewidth]{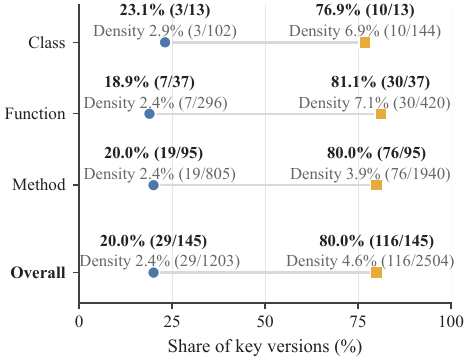}
        \caption{\texttt{R-fixed}}
        \label{fig:rq23-fix-r}
    \end{subfigure}
    \vspace{0.4em}
    \includegraphics[width=.60\linewidth]{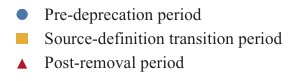}
    \caption{Lifecycle-period composition and density of key versions. Density denotes the proportion of key versions among valid release points in a stage, across the cases that have key versions.}
    \label{fig:rq23-stage-distribution}
\end{figure}

Figure~\ref{fig:rq23-stage-distribution} compares the lifecycle-period composition and key-version density for mappings with a source-definition transition period. This comparison covers 97 of the 133 \texttt{D-fixed} key versions and 145 of the 173 \texttt{R-fixed} key versions.

In \texttt{D-fixed}, 64/97 key versions (66.0\%) occur in the post-removal period, compared with 33/97 (34.0\%) in the source-definition transition period. The post-removal period also has a higher key-version density: 8.8\% (64/728) of its valid release points are key versions, compared with 2.9\% (33/1,134) in the source-definition transition period. Thus, the higher concentration remains after accounting for the different numbers of valid release points. In \texttt{R-fixed}, 116/145 key versions (80.0\%) occur in the source-definition transition period, compared with 29/145 (20.0\%) in the pre-deprecation period. The source-definition transition period also has the higher density, 4.6\% (116/2,504) versus 2.4\% (29/1,203).

\begin{figure}[pos=t]
    \centering
    \includegraphics[width=.9\linewidth]{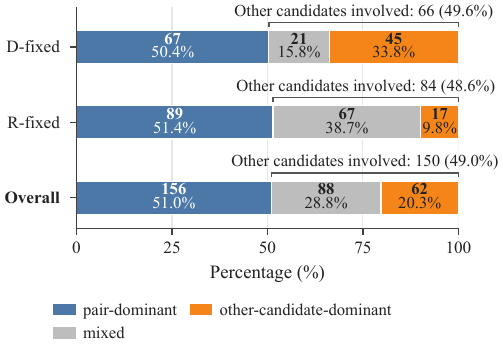}
    \caption{Ranking-change source attribution at key versions.}
    \label{fig:rq-dominant-cause}
\end{figure}

We next examine whether the key ranking changes arise mainly from the deprecated--replacement pair or from other candidates.
Figure~\ref{fig:rq-dominant-cause} separates pair-dominant, mixed, and other-candidate-dominant attributions. In \texttt{D-fixed}, the three categories account for 50.4\% (67/133), 15.8\% (21/133), and 33.8\% (45/133), respectively. In \texttt{R-fixed}, they account for 51.4\% (89/173), 38.7\% (67/173), and 9.8\% (17/173). Combining mixed and other-candidate-dominant cases, 150/306 key versions (49.0\%) involve other-candidate effects.

To examine how these attribution categories relate to changes in the surrounding candidate set, we further analyze the \texttt{D-fixed} scenario, in which candidates evolve across target releases. Table~\ref{tab:rq22-candidate-changes} reports two measures. The first is the proportion of key versions for which the absolute difference in candidate-set size between $v_i$ and $v_{i+1}$ is at least 10\% of the candidate count in $v_i$. The second, computed for each attribution category, is the number of removed, added, or source-modified candidate instances divided by the total number of candidates in the preceding releases $v_i$ of its key-version transitions.

The two measures capture different aspects of candidate-set change. A net candidate-count change of at least 10\% occurs in 93.3\% (42/45) of other-candidate-dominant key versions, compared with 7.5\% (5/67) of pair-dominant and 0.0\% (0/21) of mixed key versions. The changed-candidate ratio increases from 30.1\% for pair-dominant cases to 63.1\% for mixed cases and 72.3\% for other-candidate-dominant cases. Thus, candidate sets change most broadly when other candidates dominate the ranking change. Mixed cases also involve large candidate changes even though their net candidate counts usually remain stable.

\begin{finding}
Key versions concentrate in the post-removal period in \texttt{D-fixed} and the source-definition transition period in \texttt{R-fixed}, both in share and density. Other candidates affect 49.0\% of key versions overall. In \texttt{D-fixed}, other-candidate-dominant cases coincide with the largest candidate-set changes.
\label{finding:5}
\end{finding}

\begin{table}[width=\columnwidth,pos=t]
\centering
\caption{Candidate-set changes across ranking-change attribution categories in the \texttt{D-fixed} scenario.}
\label{tab:rq22-candidate-changes}
\setlength{\tabcolsep}{2.5pt}
\renewcommand{\arraystretch}{1.06}
\small
\begin{tabularx}{.99\columnwidth}{@{}>{\raggedright\arraybackslash}p{2.2cm}>{\centering\arraybackslash}X>{\centering\arraybackslash}X@{}}
\toprule
Attribution
& Key versions with count change $\geq 10\%$
& Changed candidates / baseline candidates \\
\midrule
Pair-dominant
& 5/67 (7.5\%)
& 24,987/82,957 (30.1\%) \\
Mixed
& 0/21 (0.0\%)
& 11,768/18,660 (63.1\%) \\
Other-candidate-dominant
& 42/45 (93.3\%)
& 13,948/19,287 (72.3\%) \\
\bottomrule
\end{tabularx}

\vspace{0.25em}
\begin{minipage}{.98\columnwidth}
\footnotesize
\textit{Note:} Count change is the absolute change in candidate-set size relative to the preceding release. Changed candidates / baseline candidates is the ratio of removed, added, or source-modified candidate instances to preceding-release candidate instances.
\end{minipage}
\end{table}

\subsection{RQ2.3: Code-Change Patterns and Ranking Effects Across Similarity Methods}

RQ2.3 examines which source-code changes are associated with the observed ranking effects and why the similarity methods respond differently. Following the screening and coding procedure in Section~\ref{sec:rq23-design}, we analyze 304 unique analysis change blocks: 85 in \texttt{D-fixed} and 219 in \texttt{R-fixed}.

Table~\ref{tab:rq23-change-patterns} reports 13 high-frequency code-change patterns covering 93 analysis change blocks. These 13 patterns are all at the function and method granularities: the 33 class-granularity blocks (11 in \texttt{D-fixed} and 22 in \texttt{R-fixed}) are assigned to code-change patterns that each occur fewer than five times, so no class-level pattern reaches the reporting threshold. Based on inspection of their source-code changes, we group these patterns into three recurring code-evolution mechanisms: continued expansion or refactoring of the replacement API, addition to the deprecated API of logic already present in the replacement, and deprecation-oriented adaptation through delegating wrappers or decorators.

\begin{table*}[width=.98\textwidth,cols=8,pos=t]
\centering
\caption{High-frequency code-change patterns, recurring code-evolution mechanisms, and ranking effects.}
\label{tab:rq23-change-patterns}
\setlength{\tabcolsep}{2.8pt}
\renewcommand{\arraystretch}{1.03}
\begin{tabular}{>{\raggedright\arraybackslash}p{2.3cm}l>{\raggedright\arraybackslash}p{3.6cm}>{\raggedright\arraybackslash}p{3.8cm}cccr}
\toprule
Scenario / attribution & Gran. & Mechanism / adaptation form & Code-change pattern & Graph & Token & Tree & Freq. \\
\midrule
\multirow{4}{*}{\shortstack[l]{\texttt{D-fixed}\\Pair-dominant}}
& Function & \multirow{4}{3.6cm}{Replacement expansion or refactoring} & Add conditional statement       & $\downarrow$ & $\downarrow$ & $\downarrow$ & 6 \\
& Method   &                         & Add conditional statement       & $\downarrow$ & $\downarrow$ & $\downarrow$ & 6 \\
& Method   &                         & Add call-assignment statement   & $\downarrow$ & $\downarrow$ & $\downarrow$ & 5 \\
& Method   &                         & Modify method implementation    & $\downarrow$ & $\downarrow$ & $\downarrow$ & 5 \\
\midrule
\shortstack[l]{\texttt{R-fixed}\\Mixed} & Method & Shared-logic addition & Add conditional statement & $\uparrow$ & $\uparrow$ & $\uparrow$ & 5 \\
\midrule
\multirow{2}{*}{\shortstack[l]{\texttt{R-fixed}\\Mixed}}
& Method   & \multirow{6}{3.6cm}{Deprecation-oriented adaptation: delegating wrapper} & Modify method implementation    & $\downarrow$ & $\downarrow$ & $\downarrow$ & 10 \\
& Method   &                         & Add return-call statement       & $\downarrow$ & ---           & $\downarrow$ & 7 \\
\multirow{4}{*}{\shortstack[l]{\texttt{R-fixed}\\Pair-dominant}}
& Function &                         & Modify function implementation  & $\downarrow$ & $\downarrow$ & $\downarrow$ & 11 \\
& Function &                         & Modify implementation block     & $\downarrow$ & $\downarrow$ & $\downarrow$ & 5 \\
& Method   &                         & Modify method implementation    & $\downarrow$ & $\downarrow$ & $\downarrow$ & 16 \\
& Method   &                         & Modify method implementation    & $\downarrow$ & ---           & $\downarrow$ & 5 \\
\midrule
\shortstack[l]{\texttt{R-fixed}\\Mixed}
& Method   & \multirow{2}{3.6cm}{Deprecation-oriented adaptation: decorator} & Add decorator                   & ---           & ---           & $\downarrow$ & 7 \\
\shortstack[l]{\texttt{R-fixed}\\Pair-dominant}
& Method   &                         & Add decorator                   & ---           & ---           & $\downarrow$   & 5 \\
\bottomrule
\end{tabular}

\vspace{0.3em}
\begin{minipage}{.96\textwidth}
\footnotesize
\textit{Note:} $\uparrow$, ---, and $\downarrow$ denote improvement, no change, and decline in the replacement API ranking. Frequency is the number of unique analysis change blocks. Delegating wrappers and deprecation decorators are two forms of deprecation-oriented adaptation. Only patterns with a frequency of at least 5 are reported.
\end{minipage}
\end{table*}

The first mechanism includes four \texttt{D-fixed} patterns and 22 analysis change blocks. These patterns occur mostly after source-definition removal and reflect continued expansion or refactoring of the replacement API. They introduce conditions, calls, or implementation structures that are absent from the fixed deprecated API, and all three methods report ranking decline for all 22 blocks.

The second mechanism includes one \texttt{R-fixed} pattern with five analysis change blocks. In \texttt{R-fixed}, the replacement remains fixed at the source-definition removal release, while the deprecated API varies across the releases preceding removal. In these blocks, later versions of the deprecated API add conditional logic that is also present in the fixed replacement. This common logic increases overlap in tokens, dependencies, and syntax, and all three methods report ranking improvement.

The third mechanism dominates the high-frequency results in \texttt{R-fixed}, accounting for 66/71 analysis change blocks (93.0\%). It includes six delegating-wrapper patterns with 54 blocks and two deprecation-decorator patterns with 12 blocks. Among the delegating-wrapper blocks, 42 produce ranking decline under all three methods, while 12 produce graph- and tree-based decline with no token-based rank change. Listing~\ref{lst:rq23-delegating-wrapper} shows an example from the more common group, in which the deprecated API's local implementation is replaced by direct delegation to the replacement and all three methods report a ranking decline.

\begin{lstlisting}[float=tbp,
    style=diffstyle,
    caption={Representative delegating-wrapper change from SciPy 1.5.4 to 1.6.0 in the \texttt{R-fixed} scenario. All three rankings decline.},
    label={lst:rq23-delegating-wrapper}
]
@@ cumtrapz: SciPy 1.5.4 -> 1.6.0 @@
 def cumtrapz(y, x=None, dx=1.0, axis=-1,
             initial=None):
-    y = np.asarray(y)
-    if x is None:
-        d = dx
-    else:
-        ...
-    res = np.cumsum(...)
-    return res
+    return cumulative_trapezoid(
+        y, x=x, dx=dx, axis=axis,
+        initial=initial
+    )
\end{lstlisting}

Deprecation-decorator additions account for the remaining 12 deprecation-oriented instances and show clearer differences among the code representations. Graph-based and token-based rankings remain unchanged in all 12 instances, whereas the tree-based ranking declines in all 12. The graph representation does not model decorators. The token representation includes decorator-related elements, but these additions do not change the replacement's position relative to other candidates in these instances. The tree representation encodes decorators as syntax nodes and therefore responds to the added surrounding structure. Listing~\ref{lst:rq23-decorator} illustrates this tree-based decline: only the surrounding decorator syntax is added, while the method body remains unchanged.

\begin{finding}
High-frequency patterns link ranking changes to code evolution. In \texttt{D-fixed}, replacement expansion or refactoring introduces structures absent from the fixed deprecated API and lowers rankings. In \texttt{R-fixed}, later deprecated-API versions may add logic already present in the fixed replacement and improve rankings. Delegating wrappers generally lower rankings; decorators leave graph and token rankings unchanged but lower tree rankings.
\label{finding:6}
\end{finding}

\begin{lstlisting}[float=t,
    style=diffstyle,
    caption={Representative deprecation-decorator addition from Matplotlib 3.3.4 to 3.4.0 in the \texttt{R-fixed} scenario. Graph and token rankings remain unchanged, while the tree ranking declines.},
    label={lst:rq23-decorator}
]
@@ get_window_title: Matplotlib 3.3.4 -> 3.4.0 @@
+@_api.deprecated(
+    "3.4",
+    alternative=("manager.get_window_title "
+                 "or GUI-specific methods"),
+)
 def get_window_title(self):
     if self.manager is not None:
         return self.manager.get_window_title()
\end{lstlisting}

%% file: sec5-rq3.tex
\section{RQ3: Post-Deprecation Lifecycle}

\subsection{RQ3.1: Prevalence of Transition Periods}

A deprecation announcement does not necessarily trigger either ending event immediately. We therefore compare how often the source-definition and original-invocation transition periods defined in Section~\ref{sec2_rq3} occur.

\begin{figure}[pos=t]
    \centering
    \includegraphics[width=.9\linewidth]{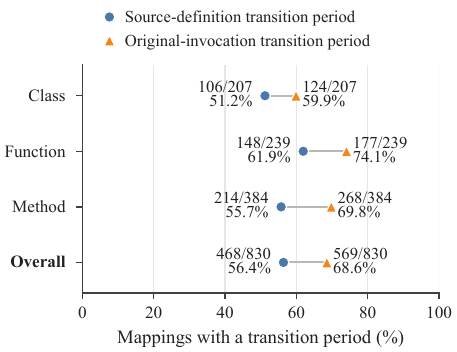}
    \caption{Proportions of mappings with source-definition and original-invocation transition periods.}
    \label{fig:rq31-transition}
\end{figure}

Figure~\ref{fig:rq31-transition} shows that a source-definition transition period exists for 468/830 mappings (56.4\%), whereas an original-invocation transition period exists for 569/830 mappings (68.6\%). The same pattern holds at all three granularities: the invocation-based proportion exceeds the source-definition-based proportion by 8.7 percentage points for classes, 12.2 for functions, and 14.1 for methods (12.2 points overall). Thus, the original invocation remains executable beyond the announcement in a larger proportion of mappings than the original source definition remains present. This overall difference shows that the two measures capture distinct lifecycle states. RQ3.2 examines the timing for individual mappings.

\begin{finding}
Post-deprecation transition periods are common: 56.4\% of mappings have a source-definition transition period and 68.6\% have an original-invocation transition period. The higher invocation-based proportion shows that post-announcement executability is more prevalent than retention of the original source definition, indicating that the two measures capture distinct lifecycle states.
\label{finding:7}
\end{finding}

\subsection{RQ3.2: Timing Differences and Underlying Mechanisms}

The difference between the two transition-period proportions suggests that source-definition removal and original-invocation failure do not always occur in the same release. We first compare their release-level timing and then examine the mechanisms in invocation-first and source-first cases.

\begin{figure}[pos=t]
    \centering
    \includegraphics[width=.9\linewidth]{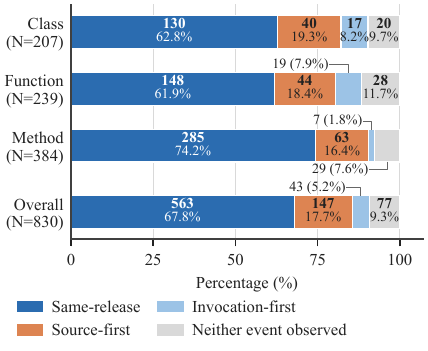}
\caption{Timing of source-definition removal and original-invocation failure.}
    \label{fig:rq32-temporal}
\end{figure}

Across all 830 mappings, Figure~\ref{fig:rq32-temporal} identifies 563 same-release cases, 147 source-first cases, 43 invocation-first cases, and 77 cases in which neither ending event is observed. Here, \textit{same-release} means that the two ending events occur in the same release, not that they occur with the deprecation announcement. Because neither event is observed for the last group, we exclude these 77 mappings from the timing comparison. Among the remaining 753 mappings, 563 (74.8\%) are same-release, while 190 (25.2\%) are source-first or invocation-first. Source-first and invocation-first also include cases in which the second event remains unobserved at the lifecycle observation cutoff. Of these 190 cases, 147 (77.4\%) are source-first and 43 (22.6\%) are invocation-first. Source-first cases outnumber invocation-first cases at every API granularity.

\begin{table*}[width=.9\textwidth,cols=4,pos=t]
\centering
\caption{Mechanisms associated with different event timing ($N=190$).}
\label{tab:rq32-mechanisms}
\setlength{\tabcolsep}{6pt}
\renewcommand{\arraystretch}{1.08}
\begin{tabularx}{\linewidth}{l>{\raggedright\arraybackslash}Xrr}
\toprule
Direction & Mechanism & $N$ & Share \\
\midrule
\multirow{4}{*}{Invocation-first}
& Deprecated access path or public alias unavailable & 20 & 46.5\% \\
& Definition retained with explicit failure logic & 9 & 20.9\% \\
& Incompatible invocation contract & 6 & 14.0\% \\
& Repository definition inconsistent with runtime or package exposure & 8 & 18.6\% \\
\midrule
\multirow{4}{*}{Source-first}
& Compatibility through explicit aliases or re-exports & 84 & 57.1\% \\
& Method compatibility through inheritance & 52 & 35.4\% \\
& Dynamic attribute resolution & 4 & 2.7\% \\
& Same-named wrapper or namespace substitution & 7 & 4.8\% \\
\bottomrule
\end{tabularx}
\end{table*}

Table~\ref{tab:rq32-mechanisms} shows that invocation-first cases arise at several points between a repository definition and a user's call. The most frequent mechanism is removal of the deprecated access path or public alias (20/43, 46.5\%). In these cases, the definition remains in the repository, but the original name is no longer exposed. Listing~\ref{lst:rq32-jax-access} shows how JAX disables the deprecated top-level binding while retaining the internal definition.

\begin{lstlisting}[float=t,
    style=diffstyle,
    caption={Invocation-first example: JAX disables the public \texttt{jax.tree\_flatten} binding from version 0.5.3 to 0.6.0.},
    label={lst:rq32-jax-access}
]
@@ jax/__init__.py: 0.5.3 -> 0.6.0 @@
-from jax._src.tree_util import (
-    tree_flatten as _deprecated_tree_flatten
-)
 _deprecations = {
-    "tree_flatten": (..., _deprecated_tree_flatten)
+    "tree_flatten": (..., None)
 }
\end{lstlisting}

The remaining invocation-first cases retain a definition whose body explicitly fails (9 cases), change the invocation contract through parameters or API type (6), or expose a runtime or released-package object that differs from the repository definition (8). Thus, the presence of a definition alone does not establish that the original invocation remains executable.

The mechanisms in source-first cases are concentrated in two categories. Explicit aliases and re-exports account for 84/147 cases (57.1\%), and inheritance accounts for 52 (35.4\%). Together, they account for 92.5\% of source-first cases. Aliases and re-exports preserve a resolution path from the deprecated API's original name or access path to a valid implementation after its original definition is removed. Listing~\ref{lst:rq32-compatibility-bindings} presents two representative implementations of this most frequent mechanism.

\begin{lstlisting}[float=t,
    style=casestyle,
    language=Python,
    caption={Source-first compatibility bindings after definition removal or migration.},
    label={lst:rq32-compatibility-bindings}
]
# SymPy 1.13.0: preserve sympy.divisor_sigma through re-export
from .functions import divisor_sigma

# Tornado 6.3.0: bind the deprecated name to its replacement
get_secure_cookie = get_signed_cookie
\end{lstlisting}

Inheritance similarly preserves a method call when a subclass definition is removed but an identically named method remains reachable through the method resolution order. The remaining 11 source-first cases use dynamic attribute resolution, same-named wrappers, or namespace substitution. Across these mechanisms, removal from the original source path does not make the original invocation fail as long as the old name or method lookup still resolves to a valid implementation.

\begin{finding}
Among mappings with at least one observed ending event, 25.2\% are source-first or invocation-first; 77.4\% of these are source-first. Aliases, re-exports, and inheritance largely explain source-first cases, whereas invocation-first cases arise from several mechanisms that disrupt access, execution, invocation contracts, or runtime exposure.
\label{finding:8}
\end{finding}

%% file: sec6-implications.tex
\section{Implications}

\subsection{Implications for Replacement API Recommendation}

\noindent \textit{\textbf{Granularity- and Competition-Aware Candidate Retrieval.}}
Replacement locality varies by API granularity, and other-candidate effects occur in 49.0\% of the analyzed key versions. These results make candidate generation part of the ranking problem rather than fixed preprocessing. When the replacement is unknown, code-similarity-based recommenders should use locality as a granularity-specific search priority rather than a hard boundary. They should prioritize the original class and module for methods, search the original module before expanding for functions, and expand across modules earlier for classes. Candidate retrieval and ranking should be evaluated together under changes to the candidate set.

\noindent \textit{\textbf{Version- and Representation-Aware Ranking.}}
Source and target releases and code representations are not neutral evaluation choices. Replacement rankings vary across releases, and the evaluated similarity methods agree on the trend in only about 60\% of cases. Code-similarity-based recommendation studies should therefore specify the release pairs used and report cross-version and cross-representation ranking stability. Recommenders could compare the top-ranked candidates across relevant release pairs and code representations. Stable rankings provide more robust recommendation evidence, whereas changes in the top candidates or disagreement among methods can flag cases that require additional evidence from changelogs, documentation, or client context.

\subsection{Implications for Automated API Migration}

\noindent \textit{\textbf{Call-Site-Specific Parameter Adaptation.}}
Definition-level parameter differences should be treated as adaptation triggers, not as evidence of client failure. After a replacement candidate is selected or confirmed, migration tools should combine these differences with actual call arguments to identify calls that may require adaptation and propose argument-level changes.

\noindent \textit{\textbf{Lifecycle and Compatibility Evidence.}}
Source-definition removal should not be used as a proxy for original-invocation failure. Lifecycle analyses and migration tools should resolve public names and access paths and account for compatibility mechanisms beyond explicit definitions. Delegating wrappers can directly reveal the implementation that receives old invocations, while aliases, re-exports, and inheritance can trace how old access paths are preserved. These relations provide concrete migration evidence for identifying and validating replacement candidates and for explaining how existing calls are redirected across releases. Library maintainers should document and test the end of support for old access paths separately from source-definition removal.

%% file: sec7-threats.tex
\section{Threats to Validity}

\noindent \textbf{\textit{Internal Validity.}}
Retrieving and interpreting deprecated API--replacement API records may introduce errors. We manually screened dedicated changelog sections and used pilot-derived keywords elsewhere. These keywords covered 99.46\% of the pilot entries but may miss unseen expressions. Two researchers independently assessed the records, resolved disagreements, and validated the mappings against source definitions. Locating definitions, tracking lifecycle events, and executing original invocations may also be affected by complex source structures or environments. We cross-checked source and release evidence and excluded failures unrelated to the target API. Finally, the qualitative coding in RQ2.3 and RQ3 may depend on researcher judgment. To mitigate this threat, one researcher coded all cases, and another reviewed every label.

\noindent \textbf{\textit{External Validity.}}
The dataset contains 830 same-library, same-granularity mappings from 33 widely used Python libraries. The findings may not generalize to smaller or less documented libraries, cross-library or cross-granularity replacements, or indirect API uses. RQ2 evaluates three widely used code-similarity representations. However, other implementations, learned representations, or additional evidence may produce different rankings. RQ3 is limited by the August 1, 2025 observation cutoff: 77 mappings had neither ending event and were excluded from the timing comparison, so their future outcomes may change the reported proportions.

\noindent \textbf{\textit{Construct Validity.}}
RQ1 derives interface differences from definitions. These differences indicate potential adaptation needs but do not establish client failure or semantic equivalence. RQ2 uses the similarity-based rank of the maintainer-specified replacement as an evaluation measure of recommendation behavior, rather than the accuracy of a complete recommendation system. Its results depend on the candidate scope, which covers 73.9\% of class, 79.5\% of function, and 99.0\% of method replacements before scenario filtering, and on the selected statistical tests and thresholds. Serial dependence in rank sequences may also affect Mann--Kendall significance. Alternative choices may change the reported trends and key versions. RQ3 uses one validated original invocation per mapping. Its failure does not imply that every argument, access path, or client context fails in the same release.

%% file: sec8-related_work.tex
\section{Related Work}

\noindent \textbf{\textit{Python API Evolution and Deprecation.}}
API evolution and deprecation studies examine API changes, deprecation practices, and client responses across language ecosystems~\citep{robbes2012api-deprecation,jezek2015java-api-breaking}. Brito et al.~\citep{brito2018api-deprecation-replacement} measured the use and evolution of replacement messages, while Sawant et al.~\citep{sawant2019deprecation-reactions} tracked how clients respond to deprecated API elements. In Python, prior work analyzes framework evolution and client effects~\citep{zhang2020python-api-evolution,zhang2021tensorflow-api-evolution}, deprecation practices~\citep{wang2020python-deprecation,zhong2025python-package-deprecation}, breaking changes under dynamic language features~\citep{du2022aexpy}, and behavioral changes caused by default arguments~\citep{montandon2025default-argument-breaking}. Our study instead uses maintainer-specified deprecated API--replacement API relationships as the unit of analysis and connects their structural, release, and lifecycle contexts.

\noindent \textbf{\textit{Replacement API Identification and Recommendation.}}
Prior approaches identify replacements from version differences and evolution evidence. Diff-CatchUp and SemDiff infer adaptations from framework changes and framework usage~\citep{xing2007diff-catchup,dagenais2008semdiff}, while AURA and HiMa recover evolution rules from call dependencies, text, and version histories~\citep{wu2010aura,meng2012hima}. REPFINDER characterizes the information sources, code changes, and mapping cardinalities of missing APIs and their replacements before using this evidence to find replacements~\citep{huang2021repfinder}. Lamothe and Shang~\citep{lamothe2018android-api-migration} showed that Android replacements may be introduced in releases other than those in which the original APIs are deprecated or removed. A recent study by Venkatakrishnan et al.~\citep{venkatakrishnan2026libshift} proposed LibShift Search, which uses code, name, and docstring embeddings to retrieve replacements for deprecated Python methods and evaluates 96 validated mappings from six libraries. These approaches infer replacements at selected version transitions. We instead use maintainer-specified mappings as evaluation targets and track how their candidate rankings change release by release as the deprecated API, replacement API, and other candidates evolve.

\noindent \textbf{\textit{Python API Migration and Compatibility Repair.}}
API migration and compatibility repair have been widely studied in other language ecosystems~\citep{zhong2010api-mapping-migration,fazzini2019api-usage-update,nielsen2021javascript-api-adaptation,kuang2025multiversion-api-migration}. At the project and environment levels, studies of deep-learning systems characterize cross-component compatibility issues and the causes of different compatible framework versions across client projects~\citep{wang2023dl-compatibility,lei2023framework-version-compatibility}. Python approaches include SOAR~\citep{ni2021soar}, MLCatchUp~\citep{haryono2021mlcatchup}, RELANCER~\citep{zhu2021relancer}, and PCART~\citep{zhang2026pcart}. For Python dependency upgrades, PCREQ combines version and code compatibility analysis to infer compatible requirements~\citep{lei2026pcreq}. The empirical study preceding MLCatchUp characterized 112 deprecated machine-learning API migrations by update operation, mapping cardinality, and context dependence. At the library level, Islam et al.~\citep{islam2024python-library-migration} characterized API mappings, code changes, and development effort across 335 Python library migrations. Recent studies examine deprecated API use in LLM-based code completion~\citep{wang2025deprecated-api-completion} and evaluate or organize large language models for library migration and deprecated-API updates~\citep{islam2025python-llm-migration,zhu2026deprecated-api-llm,kang2026pig}. These studies focus on system or dependency compatibility, replacement information, selected version transitions, migration operations, or client reactions. Our study instead jointly examines maintainer-specified relationships across API granularity, release-by-release ranking evolution, and post-deprecation source-definition and original-invocation states.

%% file: sec9-conclusion.tex
\section{Conclusion}

In this paper, we study 830 maintainer-specified deprecated API--replacement API mappings from 33 Python libraries through three research questions spanning structure, version evolution, and lifecycle. First, we find granularity-specific patterns in replacement relationships: replacement locality increases from classes to methods, while functions show the broadest parameter-interface variation. Second, we show that replacement rankings change across library versions as deprecated APIs, replacement APIs, and other candidates evolve. We also find that different code representations can produce different ranking trends. Third, we distinguish source-definition removal from original-invocation failure. We find that compatibility mechanisms can preserve original invocations after their definitions disappear, whereas changes to access paths or invocation contracts can cause them to fail before removal. Our findings characterize deprecated API--replacement API mappings as evolving migration relationships shaped by API structure, version evolution, candidate competition, and lifecycle events. The findings and dataset provide an empirical basis for evolution-aware replacement API recommendation and automated API migration.